\documentstyle[12pt]{article}
\renewcommand\theequation{\thesection.\arabic{equation}}

\begin{document}
\setlength{\unitlength}{1mm}

{\hfill   August 1996 }

{\hfill    WATPHYS-TH-96/11 }

{\hfill    hep-th/9609085} \vspace*{2cm} \\
\begin{center}
{\Large\bf Quantum scalar field on three-dimensional  
(BTZ) black hole instanton: heat kernel,
effective action and thermodynamics
}
\end{center}
\begin{center}
{\large\bf  Robert B.~Mann\footnote{e-mail: mann@avatar.uwaterloo.ca} 
and Sergey N.~Solodukhin\footnote{e-mail: sergey@avatar.uwaterloo.ca}\footnote{NATO Fellow}}
\end{center}
\begin{center}
{\it Department of Physics, University of Waterloo, Waterloo, Ontario N2L 3G1, 
Canada}  

\medskip

\end{center}
\vspace*{2cm}
\begin{abstract}

We consider the behaviour of a 
quantum scalar field on three-dimensional Euclidean
backgrounds: Anti-de Sitter space, the regular BTZ black hole instanton
and the BTZ instanton with a conical singularity at the horizon.
The corresponding heat kernel and effective action are calculated
explicitly for both rotating and non-rotating holes. The quantum 
entropy of the BTZ black hole is calculated by differentiating the effective 
action with respect to the angular deficit at the conical singularity.
The renormalization of the UV-divergent terms in the action and entropy
is considered. The structure of the UV-finite term in the quantum 
entropy is of 
particular interest. Being negligible for large outer horizon area
$A_+$ it behaves  logarithmically for small $A_+$. Such behaviour might be
important at late stages of black hole evaporation.
\end{abstract}
\begin{center}
{\it PACS number(s): 04.70.Dy, 04.62.+v}
\end{center}
\vskip 1cm

\newpage
\baselineskip=.8cm
\section{Introduction}
\setcounter{equation}0

Early interest in lower-dimensional black hole physics \cite{early}
has grown into a rich and fruitful field of research. The main motivation
for this is that the salient problems of quantum black holes, such as loss
of information and the endpoint of quantum evaporation, can be more
easily understood in some simple low-dimensional models
than directly in four dimensions \cite{1}. Several interesting
2D candidates have been explored to this end
which share many common features with their four-dimensional 
cousins \cite{2}. This is intriguing since the 
one-loop quantum effective action in two dimensions is exactly 
known, in the form of the Polyakov-Liouville term, giving rise to
the hope that the semiclassical treatment
of quantum black holes in two dimensions can be done explicitly
(see reviews \cite{1}).

The black hole in three-dimensional gravity discovered by 
Ba$\tilde{n}$ados, 
Teitelboim and Zanelli (BTZ) \cite{3} has features that are
even more realistic than its two-dimensional counterparts. 
It is similar to the Kerr black hole, being
characterized by mass $M$ and angular momentum $J$ and having an event 
horizon and
(for $J\neq 0$) an inner horizon \cite{4,rev1,rev2,rev3}.
This solution naturally appears
as the final stage  of collapsing matter \cite{5}. 
In contrast to the Kerr solution it is asymptotically Anti-de Sitter 
rather than asymptotically flat.  Geometrically, the BTZ 
black hole is obtained from 3D Anti-de Sitter (AdS$_3$)
spacetime by performing some identifications. Although
quantum field theory on curved three-dimensional manifolds
is not as well understood as in two dimensions, the large symmetry
of the BTZ geometry and its relation to AdS$_3$ allow one to
obtain some precise results when field is quantized  
on this background. The Green's function
and quantum stress tensor for the conformally coupled scalar field 
and the resultant back reaction  were 
calculated in \cite{6,7,8}.  

The possibility that black hole entropy might have
a statistical explanation remains an intriguing issue, and there has
been much recent activity towards its resolution via a variety
of approaches (for a review see \cite{9}). One such proposal is
that the Bekenstein-Hawking entropy
is completely generated by quantum fields propagating
in the black hole background. Originally it was belived that
UV-divergent quantum corrections associated with such fields
to the Bekenstein-Hawking expression play a fundamental role in the statistical
interpretation of the entropy. However, it was subsequently
realized that these divergent corrections can be associated with
those that arise from the standard UV-renormalization
of the gravitational couplings in the effective action \cite{10}, \cite{J},
\cite{SS}, \cite{FS}, \cite{SS2}, \cite{MS}. 
The idea of complete generation of the 
entropy by quantum matter in the spirit of induced gravity \cite{J} 
encountered the problem of an appropriate statistical treatment
of the entropy of non-minimally-coupled matter \cite{Kab}, \cite{FFZ}
(see, however, another realization of this idea within the superstring
paradigm \cite{D}). At the same time, it was argued in number of 
papers \cite{SS1,FWS,F1,Zas} that UV-finite quantum corrections to the
Bekenstein-Hawking
entropy might be even more important than the UV-infinite ones. They
could provide essential modifications of the thermodynamics of a hole
at late stages of the evaporation when quantum effects come to play.

Relatively little work has been done concerning 
the quantum aspects of the entropy for the BTZ black hole.
Carlip \cite{Carlip} has shown that the 
appropriate quantization of 3D gravity represented in the Chern-Simons 
form yields a set of boundary states at the horizon. These can be counted
using methods of Wess-Zumino-Witten theory. Remarkably, 
the logarithm of their number gives the classical Bekenstein-Hawking 
formula. This is the unique case of a statistical explanation of 
black hole entropy. Unfortunately, it is essentially based on features 
peculiar to three-dimensional gravity and its extension to four dimensions 
is not straightforward.
An investigation of the thermodynamics of quantum scalar fields on 
the BTZ background \cite{Ichinose} concluded 
that the divergent terms in
the entropy are not always due to the existence of the outer horizon
(i.e. the leading term in the quantum entropy is not proportional to the 
area of the 
outer horizon) and depend upon the regularization method. This
conclusion seems to be in disagreement with the  
expectations based on the study of the problem in two and 
four dimensions.

In this paper we systematically calculate
the heat kernel, effective action and quantum entropy of scalar matter
for the BTZ black hole. The relevant operator is
$(\Box+\xi /l^2)$,  where 
$\xi$ is an arbitrary constant 
and $1/l^2$ is the cosmological constant appearing in the BTZ solution.
 Since we are interested in the 
thermodynamic aspects we consider the Euclidean BTZ 
geometry with a conical singularity at the horizon as the background.
In the process of getting the heat kernel and effective action on this
singular geometry we proceed in steps, first calculating quantities 
explicitly for AdS$_3$,  then 
the regular BTZ instanton and finally the conical BTZ
instanton. The entropy is calculated by differentiating the effective 
action with respect to the angular deficit at the horizon.
It contains both UV-divergent and UV-finite terms. The analysis of the 
divergences shows that they are
explicitly renormalized by renormalization of Newton's constant in
accordance with general arguments \cite{FS}. 

We find the structure of the UV-finite terms in the entropy to be
particularly interesting. These terms, negligible
for large outer horizon area $A_+$, behave logarithmically at small
$A_+$. Hence they should become important at late stages 
of black hole evaporation.

The paper is organized as follows. In Section 2 we briefly review
the Euclidean BTZ geometry, omitting details that 
appear in earlier work \cite{3,4,rev1,rev2,rev3}. We discuss
in section 3 various forms of the metric for 3D Anti-de Sitter space 
giving expressions for the geodesic distance 
 on AdS$_3$ that are relevant for our purposes.
We solve explicitly the heat kernel equation and find the Green's function 
on AdS$_3$ as a function of the geodesic distance. 
In Section 4 we calculate explicitly the trace of
heat kernel and the effective action on the regular and singular 
Euclidean BTZ instantons. The quantum entropy is the subject of Section 5
and in Section 6 we provide some concluding remarks.

\bigskip

\section{Sketch of BTZ black hole geometry}
\setcounter{equation}0

We start with the black hole metric written in a form 
that makes it similar to the four-dimensional Kerr metric. 
Since we are interested in its thermodynamic behaviour,
we write the metric in the Euclidean form:
\begin{equation}
ds^2=f(r)d\tau^2+f^{-1}(r) dr^2+r^2(d\phi+N(r)d\tau)^2~~,
\label{1}
\end{equation}
where the metric function $f(r)$ reads
\begin{equation}
f(r)={r^2\over l^2}-{j^2\over r^2}-m={(r^2-r_+^2)(r^2+|r_-|^2)\over l^2 r^2}
\label{2}
\end{equation}
and we use the notation
\begin{equation}
r^2_+={ml^2\over 2}(1+\sqrt{1+({2j\over m l})^2})~,
~~|r_-|^2={ml^2\over 2}(\sqrt{1+({2j\over m l})^2}-1)~~
\label{3}
\end{equation}
where we note the useful identity 
$$
r_+|r_-|=jl~~
$$
for future reference.
The function $N(r)$ in (\ref{1}) is 
\begin{equation}
N(r)=-{j\over r^2}~~.
\label{5}
\end{equation}

In order to transform the metric (\ref{1}) to Lorentzian  singnature we
need to make the transformation: 
$\tau\rightarrow \imath ~t$, $j\rightarrow -\imath ~j$.
Then we have that
\begin{eqnarray}
&&r_+\rightarrow r^L_+=\sqrt{ml^2\over 2}~\left(1+\sqrt{1-({2j\over m l})^2}~\right)^{1/2}~~, \nonumber \\
&&|r_-|\rightarrow \imath ~r_-^L=\sqrt{ml^2\over 2}~\left(1-\sqrt{1-({2j\over m l})^2}~\right)^{1/2}~~,
\label{?}
\end{eqnarray}
where $r^L_+$ and $r^L_-$ are the values in the Lorentzian space-time. These 
are the  respective radii of the outer and inner horizons  of the 
Lorentzian  black hole in $(2+1)$ dimensions. Therefore
we must always apply the transformation (\ref{?}) after carrying out
all calculations in the Euclidean geometry  in order to
obtain the result for the Lorentzian  black hole.
The Lorentzian version of the metric (\ref{1}) describes a black hole with mass $m$ and angular 
momentum $J=2j$ \cite{3}, \cite{4}.
Introducing 
\begin{equation}
\beta_H\equiv {2\over f'(r_+)}={r_+ l^2 \over r^2_++|r_-|^2}
\label{6}
\end{equation}
we find that in the $(\tau , r)$ sector of the metric (\ref{1}) there is 
no conical singularity at the horizon if the Euclidean time $\tau$ 
is periodic with period $2\pi\beta_H$. 
The quantity $T_H=(2\pi\beta_H)^{-1}$ is the 
Hawking  temperature of the hole.

The horizon $\Sigma$ is a one-dimensional space with metric
$
ds^2_{\Sigma}=l^2d\psi^2~~,
$
where $\psi={r_+\over l} \phi-{|r_-|\over l^2}\tau$ is a natural coordinate on the horizon.

Looking at the metric (\ref{1}) one can conclude that there is no  
constraint on the periodicity of the ``angle'' variable $\phi$ (or $\psi$).
This is in contrast to the four-dimensional black hole, for which
the angle $\phi$ in the spherical line 
element $(d\theta^2+\sin^2\theta d\phi^2)$ varies between the limits
$0\leq \phi\leq 2\pi$ in order to avoid the appearance of the 
conical singularities at the poles of the sphere. 
However, following tradition we will assume that the metric
(\ref{1}) is periodic in $\phi$, with limits
$0\leq \phi\leq 2\pi$. This means that $\Sigma$ is a circle with 
length (``area'') $A_+=2\pi r_+$.

There are a number of other useful forms for the metric 
(\ref{1}). It is very important for our considerations that (\ref{1}) 
is obtained from the metric of three-dimensional Anti-de Sitter 
space by making certain identifications along the trajectories of
its Killing vectors. In order to find the appropriate 
metric for the 3D Anti-de Sitter space (denoted below by
 $H_3$) 
we consider a four-dimensional flat space with 
metric
\begin{equation}
ds^2=dX^2_1-dT^2_1+dX^2_2+dT^2_2~~.
\label{7}
\end{equation}
AdS$_3$ ($H_3$) is defined as a subspace defined by the equation
\begin{equation}
X^2_1-T^2_1+X^2_2+T^2_2=-l^2~~.
\label{8}
\end{equation}
Introducing the coordinates $(\psi , \theta , \chi )$ 
\begin{eqnarray}
&&X_1={l\over \cos \chi}\sinh \psi~,~~ T_1={l\over \cos \chi}\cosh \psi \nonumber \\
&&X_2=l~ \tan \chi \cos \theta ~,~~T_2=l~ \tan \chi \sin \theta 
\label{9}
\end{eqnarray}
the metric on $H_3$ reads
\begin{equation}
ds^2_{H_3}={l^2\over \cos^2 \chi }(d\psi^2+d\chi^2+\sin^2 \chi d\theta^2 )~~.
\label{10}
\end{equation}
It is easy to see that under the 
coordinate transformation
\begin{eqnarray}
&&\psi={r_+\over l} \phi-{|r_-|\over l^2}\tau~, ~~\theta={r_+\over l} \tau
+{|r_-|\over l^2}\phi \nonumber \\
&&\cos \chi =({r^2_++|r_-|^2 \over r^2+|r_-|^2})^{1/2}~~
\label{11}
\end{eqnarray}
the metric (\ref{1}) coincides with (\ref{10}).
In the next section we will derive  a few other forms of the metric 
on $H_3$ which are useful in the context of calculation of the heat 
kernel and Green's function on $H_3$.

The BTZ black hole ($B_3$) described by the metric (\ref{1}) 
is obtained from  AdS$_3$
with metric (\ref{10}) by making the
following identifications:

\noindent
$i).$ $(\psi , \theta , \chi ) \rightarrow (\psi , \theta+2\pi , \chi )$.
This means that $(\phi , \tau , r) \rightarrow (\phi+\Phi , \tau+T^{-1}_H , 
r)$, where $\Phi=T^{-1}_Hj r^{-2}_+$. 

\noindent
$ii).$ $(\psi , \theta , \chi ) \rightarrow (\psi+2\pi {r_+\over l} , \theta+2\pi
{|r_-|\over l} , \chi )$, which is the analog 
of $(\phi , \tau , r) \rightarrow (\phi+2\pi , \tau , r)$.

The coordinate $\chi$ is the analog of the radial coordinate $r$. 
It has the range $0\leq \chi\leq {\pi\over 2}$. 
The point $\chi=0$ is the horizon ($r=r_+$) while $\chi={\pi\over 2}$
lies at infinity. Geometrically, $i)$ means that there is no  conical
singularity at the horizon, which is easily seen from (\ref{10}).
A section of BTZ black hole at   fixed $\chi$ is illustrated in Fig.1
for the non-rotating  ($|r_-|=0$) and   rotating cases. 
The opposite sides of the quadrangle in Fig.1 are identified. 
Therefore, the whole section looks like a torus. In the rotating
case the torus is deformed with deformation parameter $\gamma$, where 
$\tan \gamma={r_+\over |r_-|}$. 
The whole  space $B_3$ is a region between two semispheres with  
$R=\exp( \psi )$ being radius, $\chi$ playing the role of azimuthal angle 
and $\theta$ being the orbital angle. The boundaries of the region are 
identified  according to $ii)$.  

\bigskip

\section{3D Anti-de Sitter space: geometry, heat kernel and Green's function}
\setcounter{equation}0

{\bf 3.1 Metric on $H_3$}

3D Anti-de Sitter space ($H_3$) is defined as a 3-dimensional subspace of 
the flat four-dimensional space-time with metric
\begin{equation}
ds^2=dX_1^2-dT^2_1+dX^2_2+dT^2_2
\label{2.1}
\end{equation}
satisfying the constraint
\begin{equation}
X_1^2-T^2_1+X^2_2+T^2_2=-l^2~~.
\label{2.2}
\end{equation}
We are interested in AdS$_3$, which has Euclidean signature. This is
easily done by appropriately choosing the signature in (\ref{2.1}), 
(\ref{2.2}). The induced metric has a number of different 
representations depending on the choice
of the coordinates on AdS$_3$. Below we consider two such choices.

{\bf A.} Resolve  equation (\ref{2.2}) as follows:
\begin{eqnarray}
&&X_1=l \cosh \rho \sinh \psi~,~~T _1=l \cosh \rho \cosh \psi \nonumber \\
&&X_2=l \sinh \rho \cos \theta~,~~T _1=l \sinh \rho \sin \theta~~.
\label{2.3}
\end{eqnarray}
The variables $(\rho , \psi , \theta )$ can be considered as coordinates 
on $H_3$. They are closely related to the system $(\chi , \psi , \theta )$
via the transformation $\cos \chi=\cosh^{-1}\rho$.
Note that the section of $H_3$ corresponding to a fixed $\rho$ is a 
two-dimensional torus. The induced metric then
takes the following form:
\begin{equation}
ds^2_{H_3}=l^2 \left( d\rho^2+\cosh^2\rho d\psi^2+\sinh^2\rho d\theta^2 \right)
~~.
\label{2.4}
\end{equation}
The  BTZ black hole metric is then obtained from (\ref{2.4}) by making the
identifications
$\theta\rightarrow \theta +2\pi$ and $\psi\rightarrow \psi+2\pi{r_+\over l}$,
$\theta\rightarrow\theta+2\pi{|r_-|\over l}$.

{\bf B.} Another way to resolve the constraint (\ref{2.2}) is 
by employing the transformation
\begin{eqnarray}
&&X_1=l\sinh (\sigma /l) \cos \lambda ~,~~T_1=l \cosh (\sigma /l) \nonumber \\
&&X_2=l\sinh (\sigma /l) \sin\lambda\sin\phi~,~~T_2=l \sinh (\sigma /l) \sin\lambda\cos\phi~~.
\label{2.5}
\end{eqnarray}
The section $\sigma=const$ of $H_3$ is a two-dimensional sphere. The 
induced metric in the coordinates $(\sigma , \lambda , \phi )$ 
takes the form
\begin{equation}
ds^2_{H^3}= d\sigma^2+l^2\sinh^2(\sigma /l) (d\lambda^2+\sin^2\lambda d\phi^2)
\label{2.6}
\end{equation}
from which one can easily see that $H_3$ is a 
hyperbolic version of the metric on the 3-sphere
\begin{equation}
ds^2_{S^3}= d\sigma^2+l^2\sin^2(\sigma /l) (d\lambda^2+\sin^2\lambda d\phi^2)~~,
\label{2.7}
\end{equation}
allowing us to making use of our experience with  the 3-sphere 
in understanding the geometry of $H_3$.

\bigskip

{\bf 3.2 Geodesic distance on $H_3$ }

An important fact equally applicable both to $S_3$ and $H_3$ is the 
following. Consider two different points on $S_3$ ($H_3$). 
Then we can choose the coordinate system
$(\sigma , \lambda , \phi )$ such that one of the points lies at the origin
($\sigma=0$) and the other point lies on the radius 
$(\sigma , \lambda=0 , \phi )$.
This radial trajectory joining the two points is a geodesic. 
Moreover, the geodesic distance between these two points coincides with 
$\sigma$. More generally, for the
metric (\ref{2.6}), (\ref{2.7}) the geodesic distance between two 
points with equal values of $\lambda$ and $\phi$ 
($\lambda =\lambda '~,~~\phi=\phi '$)  
is given by $|\sigma-\sigma '|=\Delta\sigma$.

In order to find the geodesic distance in the coordinate system 
$(\rho , \psi , \theta)$ (\ref{2.3}) consider the following trick.  
The two points $M$ and $M'$ in the
embedding four-dimensional space determine the vectors
${\bf a}$ and ${\bf a'}$ starting from the origin:
\begin{eqnarray}
&&{\bf a}=l \cosh \rho \sinh \psi ~{\bf x_1}+l\cosh\rho \cosh \psi ~{\bf t_1}
+l\sinh\rho \cos\theta ~{\bf x_2} +l\sinh\rho \sin \theta ~ {\bf t_2}
\nonumber \\
&&{\bf a'}=l \cosh \rho ' \sinh \psi ' ~{\bf x_1}+l\cosh\rho ' \cosh \psi ' ~{\bf t_1}
+l\sinh\rho ' \cos\theta ' ~ {\bf x_2} +l\sinh\rho ' \sin \theta ' ~ {\bf t_2}~~,
 \nonumber \\
&&
\label{2.8}
\end{eqnarray}
where $({\bf t_1}, {\bf x_1}, {\bf t_2}, {\bf x_2 })$ is an
 orthonormal basis of vectors in
the space (\ref{2.1}):
\begin{equation}
-({\bf t_1 } , {\bf t_1 })=({\bf x_1} , {\bf x_1 })
=({\bf t_2 } , {\bf t_2 })=
({\bf x_2 } , {\bf x_2 })=1~~.
\label{2.9}
\end{equation}

For the scalar product of $\bf a$ and $\bf a '$ we have
\begin{equation}
({\bf a }, {\bf a'})=l^2 \left(-\cosh^2
\rho \cosh \Delta \psi+\sinh^2\rho \cos\Delta \theta \right) ~~,
\label{2.10}
\end{equation}
where $\Delta\psi=\psi-\psi '~,~~\Delta\theta=\theta-\theta '$
and for simplicity we assumed that $\rho=\rho '$.
The scalar product $({\bf a }, {\bf a'})$ is invariant quantity not dependent on a
concrete choice of coordinates. Therefore, we can calculate it
in the coordinate system ($\sigma , \lambda , \phi$). 
In this system we have
\begin{eqnarray}
&&{\bf a}=l\cosh (\sigma /l) ~{\bf t_1'}+l\cosh (\sigma /l) ~{\bf x_1'}\nonumber \\
&&{\bf a'}=l\cosh (\sigma '/l) '~{\bf t_1'}+
l\cosh (\sigma ' /l)  ~{\bf x_1'}~~.
\label{2.11}
\end{eqnarray}
The new basis $({\bf t_1'}, {\bf x_1'}, {\bf t_2'}, {\bf x_2' })$
is obtained from the old basis $({\bf t_1}, {\bf x_1}, {\bf t_2}, {\bf x_2 })$
by some orhogonal rotation. Therefore, it satisfies the same 
identities (\ref{2.9}). In new basis we have
\begin{equation}
({\bf a }, {\bf a'})=-l^2\cosh {\Delta \sigma \over l}~~.
\label{2.12}
\end{equation}
As we explained above $\Delta\sigma$ is the geodesic distance between $M$ 
and $M'$. Equating (\ref{2.10}) and (\ref{2.12}) we finally obtain 
the expression for the geodesic distance in
terms of the coordinates $(\rho , \psi , \theta )$:
\begin{equation}
\cosh {\Delta \sigma \over l}=\cosh^2 \rho \cosh \Delta \psi-\sinh^2\rho \cos\Delta
\theta
\label{2.13}
\end{equation}
or alternatively, after some short manipulations 
\begin{equation}
\sinh^2{\Delta\sigma \over 2l}=\cosh^2\rho\sinh^2{\Delta\psi\over 2}+\sinh^2\rho \sin^2{\Delta\theta \over 2}~~.
\label{2.14}
\end{equation}
For small $\rho <<1$ and $\Delta \psi <<1$ from (\ref{2.14}) we get
\begin{equation}
\Delta\sigma ^2=l^2 \left( \Delta \psi^2+4\rho^2\sin^2{\Delta \theta \over 2}
\right)
\label{2.15}
\end{equation}
what coincides with the result for  3D flat space in cylindrical coordinates.

Note, that $\Delta \sigma$ in (\ref{2.13}), (\ref{2.14}) is the intrinsic
geodesic distance on $H_3$. It is worth comparing with the 
chordal four-dimensional distance $\Sigma$ between the points $M$ and $M'$ 
measured in the imbedding 4D space. In the coordinate system 
(\ref{2.3}) we obtain
\begin{eqnarray}
&&\Sigma^2\equiv \sum (X-X')^2=l^2 ( \cosh^2\rho (\sinh \psi -\sinh \psi ')^2
-\cosh^2\rho (\cosh \psi-\cosh \psi ')^2\nonumber \\
&&+\sinh^2\rho (\cos\theta-\cos\theta ')^2
+\sinh^2\rho (\sin \theta -\sin \theta ')^2 )~~.
\label{2.16}
\end{eqnarray}
After simplification we obtain
\begin{equation}
\Sigma^2=4l^2\sinh^2{\Delta \sigma \over 2l}~~.
\label{2.17}
\end{equation}

Consider now the point $M''$ which is antipodal to the point $M'$. It is 
obtained from $M'$ by antipodal transformation $X'\rightarrow -X'$ 
(in the coordinates $(\chi , \theta , \psi )$ the antipode has 
coordinates $(\pi-\chi , \theta , \psi )$).
The point $M''$ lies in the lower ``semisphere'' of the space $H_3$.
For some applications we will need the chordal distance 
$\hat{\Sigma}$ bewteen points $M$ and $M''$:
$\hat{\Sigma}^2=\sum (X+X')^2$, where we find
\begin{equation}
\hat{\Sigma}^2=-4l^2\cosh^2{\Delta\sigma \over 2l}~~.
\label{2.18}
\end{equation}
Here $\Delta\sigma$ is the geodesic distance between $M$ and $M'$.

\bigskip

{\bf 3.3 Heat kernel and Green's function}

Consider on $H_3$ the heat kernel equation
\begin{eqnarray}
&&(\partial_s-\Box-\xi /l^2) K(x,x',s)=0 \nonumber \\
&&K(x,x',s=0)=\delta (x,x')~~,
\label{2.19}
\end{eqnarray}
where $s$ is a proper time variable. 
The operator $(\Box +\xi /l^2)$ on $H_3$ or $B_3$ can be equivalently 
represented in the form of non-minimal coupling $(\Box-{\xi\over 6}R)$. 
For $\xi={3\over 4}$ this operator would be conformal invariant. 
This equivalence, however, is no longer valid for the space $B_3^\alpha$ 
which has a conical singularity. This is because the scalar curvature on 
a conical space has a $\delta$-function-like contribution
due to a singularity that is additional to the regular value of the 
curvature.
The $\delta$-function in the operator $(\Box -{\xi \over 6}R)$ 
has been shown \cite{SS2} to non-trivially modify the regular heat kernel.
In order to avoid the problem of dealing with this perculiarity 
we will not make use of this form of the operator and will treat the
term $\xi /l^2$ as just a constant that is unrelated to
the curvature of space-time.

The function
$K(x,x',s)$ satisfying (\ref{2.19}) can be found as some function of the 
geodesic distance $\sigma$ between the points $x$ and $x'$. The simplest 
way to do this is to use the coordinate system
$(\sigma , \lambda ,\phi )$ with the metric (\ref{2.6}) when 
both points lie on the radius: 
$\lambda=\lambda ', \phi=\phi '$. Then the Laplace 
operator $\Box=\nabla^{\mu} \nabla_{\mu}$ has  only the ``radial'' part:
\begin{equation}
\Box={1\over l^2\sinh^2 {\sigma\over l}} \partial_{\sigma}\sinh^2 ({\sigma\over l})
\partial_{\sigma}= {1\over l^2\sinh {\sigma\over l}} \partial^2_{\sigma}\sinh 
({\sigma\over l}) -l^{-2}~~.
\label{2.20}
\end{equation}
Equation (\ref{2.19}) is then easily solved and the solution takes the 
form
\begin{equation}
K_{H_3}(\sigma , s)={1\over (4\pi s)^{3/2}}{\sigma /l \over \sinh (\sigma / l)}
e^{-{\sigma^2\over 4s}-\mu{s\over l^2}}~~,
\label{2.21}
\end{equation}
where $\mu=1-\xi$. 

In the conformal case we have $\xi=3/4$ 
and $\mu=1/4$. The heat kernel (\ref{2.21}) was first found by 
Dowker and Critchley \cite{Dow-Crit} for $S_3$
(for which $\sinh (\sigma /l)$ is replaced by $\sin (\sigma /l)$)
and then was extended to the hyperbolic space $H_3$ by Camporesi 
\cite{Camporesi}.

Knowing the heat kernel function $K(\sigma , s)$ we can find the 
Green's function
$G(x, x')$ as follows
$$
G(x,x')=\int_{0}^{+\infty}ds~ K(x,x',s)~~.
$$
Applying this to  the heat kernel (\ref{2.21}) and  using the integral
\begin{equation}
\int_0^{\infty}{ds\over s^{3/2}}e^{-bs-{a^2\over 4s}}={2\sqrt{\pi}\over a}
\left(\cosh (
\sqrt{b}a)-\sinh ( \sqrt{b}a)\right)
\label{*}
\end{equation}
the Green's function on $H_3$ reads
\begin{equation}
G_{H_3}(x,x')={1\over 4\pi}{1\over l \sinh ({\sigma \over l})}\left(\cosh (\sqrt{\mu}{\sigma
\over l})-\sinh (\sqrt{\mu}{\sigma
\over l}) \right)~~,
\label{2.22}
\end{equation}
where $\sigma$ is the instrinsic geodesic distance on $H_3$ between $x$ and 
$x'$. It is important to observe that the function $G_{H_3}(x,x')$ vanishes
when $\cosh (\sqrt{\mu}{\sigma
\over l})=\sinh (\sqrt{\mu}{\sigma
\over l})$. This happens when $\sigma (x, x')=\infty$, i.e. one of the 
points lies on the equator $(\chi={\pi \over 2})$. 
This fact is important in view of the arguments of \cite{IS} that the 
correct quantization on a non-globally hyperbolic space, like AdS$_3$,
 requires the 
fixing of some boundary condition for a quantum field at infinity.
The  Green's function (\ref{2.22}) constructed by means of the heat 
kernel (\ref{2.21}) automatically satisfies the Dirichlet boundary 
condition and thus provides  for us the correct quantization on $H_3$. 
To our knowledge, the form (\ref{2.22})
of the Green's function on $H_3$ is not known in the current literature.

A special case occurs when $\xi=3/4$ and $\mu=1/4$, for which
the operator
$(\Box+\xi /l^2 )\equiv (\Box-{1\over 8}R)$ is  conformally 
invariant. In this case we get 
\begin{equation}
G_{H_3}={1\over 4\pi} \left( {1\over 2l\sinh {\sigma \over 2l}}-
{1\over 2l\cosh {\sigma \over 2l}}\right)
\label{2.23}
\end{equation}
for the Green's function. 
Using (\ref{2.16}) and (\ref{2.17}) we observe that (\ref{2.23}) has a
nice form in terms of the chordal distance in the imbedding space:
\begin{equation}
G_{H_3}(x,x')={1\over 4\pi} \left({1\over |\Sigma|}-{1\over |\hat{\Sigma}|}
\right)={1\over 4\pi} \left({1\over |X-X'|}-{1\over |X+X'|}
\right)~~.
\label{2.24}
\end{equation}
The Green's function for the conformal case in the form (\ref{2.24}) 
was reported by Steif \cite{6}.

\bigskip

\section{Heat kernel on the Euclidean BTZ instanton}
\setcounter{equation}0
\medskip

{\bf 4.1 Regular BTZ instanton} 

As was explained in Section 2 the regular Euclidean 
BTZ instanton ($B_3$) may be obtained from $H_3$
by a combination of identifications which in the coordinates
$(\rho , \theta , \psi )$ are

$i).~~\theta\rightarrow \theta+2\pi 
$

$
ii). ~~\theta\rightarrow \theta+2\pi {|r_-|\over l}~,~~\psi\rightarrow\psi+
2\pi{r_+\over l}
$

Therefore, the heat kernel $K_{B_3}$
on the BTZ instanton $B_3$ is constructed via the heat kernel $K_{H_3}$ on $H_3$ as infinite sum over images
\begin{equation}
K_{B_3}(x,x',s)=\sum_{n=-\infty}^{+\infty} K_{H_3}(\rho ,~ \rho ',~
\psi-\psi '+2\pi{r_+\over l}n,~\theta-\theta '+2\pi{|r_-|\over l}n)~~.
\label{3.1}
\end{equation}
Using the path integral representation of heat kernel we would say that
the $n=0$ term in (\ref{3.1}) is due to the direct way of connecting
points $x$ and $x'$ in the path integral. On the other hand, 
the $n\neq 0$ terms are due to uncontractible winding paths that 
go $n$ times around the circle. Note that $K_{H_3}$ automatically has 
the periodicity given in $i)$. 
Therefore the sum over images in (\ref{3.1}) provides us with the 
periodicity $ii)$. 
Assuming that $\rho=\rho '$ it can be represented in the form
\begin{eqnarray}
&&K_{B_3}=\sum_{n=-\infty}^{\infty}K_{H_3}(\sigma_n,s)~,\nonumber \\
&&\cosh {\sigma_n\over l}=\cosh^2\rho\cosh\Delta\psi_n-\sinh^2\rho
\cos \Delta \theta_n~, \nonumber \\
&&\Delta\psi_n=\psi-\psi '+2\pi{r_+\over l}n~,~~\Delta\theta_n=\theta-\theta '+2\pi{|r_-|\over l}n~~,
\label{3.2}
\end{eqnarray}
where $K_{H_3}(\sigma , s)$ takes the form (\ref{2.20}).

For the further applications consider the integral
\begin{equation}
Tr_{w}K_{B_3}\equiv \int_{B_3} K_{B_3}(\rho=\rho ', 
\psi=\psi ', \theta=\theta '+w) ~d\mu_x~~,
\label{3.3}
\end{equation}
where $d\mu_x=l^3\cosh\rho\sinh\rho d\rho 
\theta d \psi$ is the measure on $B_3$.
Note that volume of $B_3$
$$V_{B_3}=\int_{B_3}d\mu_x=l^3\int_0^{2\pi}d\theta
\int_0^{2\pi r_+\over l}d\psi\int_0^{+\infty}\cosh\rho\sinh\rho d \rho
$$
is infinite and so does not depend on $|r_-|$.
This is just a simple consequence of the geometrical
fact that the two quadrangles in Fig.1 have the same area.

The integration in (\ref{3.3}) can be easily performed if  
for a fixed $n$ we change the variable 
$\rho\rightarrow \bar{\sigma}_n=\sigma_n/l$  (see Eqs.(\ref{3.2}),
(\ref{2.13})) with the corresponding change of integration measure
$$
\cosh \rho \sinh \rho d\rho={1\over 2}{\sinh\bar{\sigma}_n d\bar{\sigma}_n
\over (\cosh\Delta\psi_n-\cos\Delta\theta_n )}=
{1\over 4}{\sinh\bar{\sigma}_n d\bar{\sigma}_n
\over (\sinh^2{\Delta\psi_n\over 2}+\sin^2{\Delta\theta_n \over 2})}~~.  
$$
Then after integration Eq.(\ref{3.3}) reads
\begin{eqnarray}
&&Tr_w K_{B_3}= V_w~{e^{-\mu\bar{s}}\over (4\pi \bar{s})^{3/2}}
+(2\pi)~({2\pi r_+\over l})~{e^{-\mu\bar{s}}\over (4\pi \bar{s})^{3/2}}~
\bar{s}~\sum_{n=1}^{\infty}~{e^{-{\Delta\psi^2_n\over 4\bar{s}}}\over (\sinh^2{\Delta\psi_n\over 2}+\sin^2{\Delta\theta_n \over 2})}~~,\nonumber \\
&&V_w= 
 \left\{
\begin{array}{ll}
  {V_{B_3}\over l^3}           & {\rm if\ } w=0 ~~, \\
(2\pi)({2\pi r_+\over l}) {1\over \sin^2{w\over 2}} {\bar{s}\over 2}  & {\rm if\ } w\neq 0 ~~,
\end{array}
\right.
\label{3.4}
\end{eqnarray}
where we defined $\bar{s}=s/l^2~,~~\Delta\psi_n={2\pi r_+\over l}n~,~~
\Delta\theta_n=w+{2\pi|r_-|\over l}n$.

The knowledge of the heat kernel allows us to calculate the effective action on $B_3$:
\begin{eqnarray}
&& W_{eff}[B_3]=-{1\over 2}\int_{\epsilon^2}^{\infty}{ds\over s} Tr_{w=0}K_{B_3}
\nonumber \\
&&=W_{div}[B_3]-\sum_{n=1}^{\infty}{1\over 4n}~{e^{-\sqrt{\mu}\bar{A}_+n}
\over (\sinh^2{\bar{A}_+n\over 2}+\sin^2{|\bar{A}_-|n\over 2})}~~,
\label{3.5}
\end{eqnarray}
where  $\bar{A}_+=A_+/l$ and $|\bar{A}_-|=|A_-|/l$ and the divergent part
of the action takes the form
\begin{eqnarray}
&&W_{div}[B_3]=-{1\over 2}{1\over (4\pi)^{3/2}}V_{B_3}\int_{\epsilon^2}^{\infty}
{ds\over s^{5/2}}e^{-\mu s}\nonumber \\
&&=-{1\over (4\pi)^{3/2}}V_{B_3}~ ({1\over 3\epsilon^3}-{
\mu^2\over \epsilon}+{2\over 3}\mu^{3/2}\sqrt{\pi}+O(\epsilon ))~~,
\label{3.5'}
\end{eqnarray}
where we used (\ref{*}) to carry out the integration over $s$
in (\ref{3.5}).

Remarkably, the expression (\ref{3.5}) is invariant under transformation:
$|\bar{A}_-|\rightarrow |\bar{A}_-|+2\pi k$. As discussed in \cite{rev2}
this is a consequence of the invariance of $B_3$ under large 
diffeomorphisms  corresponding to Dehn twists: the identifications $i)$ 
and $ii)$ determining the geometry of $B_3$ are 
unchanged if we replace 
$r_+\rightarrow r_+~,~~|r_-|\rightarrow |r_-|+kl$ for any integer $k$. 
This invariance appears only for the Euclidean black hole and disappears 
when we make the Lorentzian continuation (see discussion below).

The first quantum correction to the action due to quantization of the
three-dimensional gravity
itself was discussed in  \cite{rev2}. In this case the correction was
shown to be determined by only 
quantity $2\pi (\sinh^2 {\bar{A}_+\over 2}+\sin^2{|\bar{A}_|\over 2})$ 
related with holonomies of the BTZ instanton. 

\bigskip

{\bf 4.2 BTZ instanton with Conical Singularity}
 
The conical BTZ instanton $(B_3^{\alpha})$ is obtained 
from $H_3$ by the replacing the identification $i)$ as follows:

$i').~\theta\rightarrow \theta +2\pi\alpha
$

\noindent
and not changing the identification $ii)$. For $\alpha \neq 1$ the space 
$B_3^\alpha$ has a conical singularity at the horizon ($\rho=0$).
The heat kernel on $B^{\alpha}_3$ is constructed via the heat kernel 
on the regular instanton $B_3$ by means of the Sommerfeld formula \cite{Som},
\cite{Dow}:
\begin{equation}
K_{B_3^{\alpha}}(x,x',s)=K_{B_3}(x,x',s)+{1\over 4\pi\alpha}
\int_{\Gamma}\cot {w\over 2\alpha}~~K_{B_3}(\theta-\theta '+w,s)~~dw~~,
\label{3.6}
\end{equation}
where $K_{B_3}$ is the heat kernel (\ref{3.1}). The contour $\Gamma$ in 
(\ref{3.6}) consists of two vertical
lines, going from $(-\pi+\imath \infty )$ to $(-\pi-\imath \infty )$
and from $(\pi-\imath \infty )$ to $(\pi+\imath \infty )$ and
intersecting the real axis between the poles of the $\cot {w\over 2\alpha}$:
$-2\pi\alpha,~0$ and $0,~+2\pi\alpha$ respectively. 
For $\alpha=1$ the integrand in (\ref{3.6}) is 
a $2\pi$-periodic function and the contributions from these two vertical 
lines (at a fixed distance $2\pi$ along the real axis) 
cancel each other.

Applying (\ref{3.6}) to the heat kernel (\ref{3.4}) on $B_3$ we get 
\begin{eqnarray}
&&TrK_{B^{\alpha}_3}=TrK_{B_3}+(2\pi\alpha)~({2\pi r_+\over l})~{e^{-\mu\bar{s}}\over (4\pi \bar{s})^{3/2}}~{\bar{s}\over 2}~[~{\imath\over 4\pi\alpha}
\int_{\Gamma}{\cot {w\over 2\alpha}~dw\over \sin^2{w\over 2}}\nonumber \\
&&+
\sum_{n=1}^{\infty}e^{-{\Delta\psi^2_n\over 4\bar{s}}}{\imath\over 4\pi\alpha}
\int_{\Gamma}{\cot {w\over 2\alpha}~dw\over \sinh^2{\Delta\psi_n\over 2}+\sin^2
({w\over 2}+{\pi |r_-|\over l}n)}~]~~.
\label{3.7}
\end{eqnarray}
for the trace of  the heat kernel on $B_3^\alpha$.
Note, that the first term comes from the $n=0$ 
term (the direct paths) in the sum (\ref{3.1}),
(\ref{3.2})
while the other one corresponds to $n\neq 0$ (winding paths). 
Only the $n=0$ term leads
to appearance of UV divergences (if $s\rightarrow 0$). The 
term due to winding paths ($n\neq 0$)
 is regular in the limit $s\rightarrow 0$ due 
to the factor $e^{-{\Delta\psi^2_n \over 4\bar{s}}}$.

To analyze (\ref{3.7}) we shall consider the rotating and non-rotating
cases separately.

\noindent
{\bf Non-rotating black hole ($J=0,~|r_-|=0$)}

For this case the contour integrals in (\ref{3.7})
are calculated as follows (see (\ref{A1}), (\ref{A2}))
\begin{equation}
{\imath\over 4\pi\alpha}
\int_{\Gamma}{\cot {w\over 2\alpha}~dw\over \sin^2{w\over 2}}={1\over 3}({1
\over \alpha^2}-1)\equiv 2c_2(\alpha )~~,
\label{3.8}
\end{equation}
\begin{equation}
{\imath\over 4\pi\alpha}
\int_{\Gamma}{\cot {w\over 2\alpha}~dw\over \sinh^2{\Delta\psi_n\over 2}+
\sin^2{w\over 2}}={1\over \sinh^2{\Delta\psi_n\over 2}}\left( {1\over \alpha}
{\tanh {\Delta\psi_n\over 2}\over \tanh {\Delta\psi_n\over 2\alpha}}-1
\right)
\label{3.9}
\end{equation}
Therefore, taking into account that $Tr K_{B_3}$ is given by (\ref{3.4}) multiplied 
by $\alpha$ we get for the trace  (\ref{3.7}):
\begin{eqnarray}
&&Tr K_{B_3^{\alpha}}=\left( {V_{B_3^{\alpha}}\over l^3}+{A_+\over l}(2\pi\alpha )
c_2(\alpha )~\bar{s}~ \right){e^{-\mu\bar{s}}\over (4\pi \bar{s})^{3/2}}
\nonumber \\
&&+2\pi ~ {e^{-\mu\bar{s}}\over (4\pi \bar{s})^{3/2}}~{A_+\over l}~\bar{s}~
\sum_{n=1}^{\infty}~{\tanh {\Delta\psi_n\over 2}\over \tanh {\Delta\psi_n\over 2\alpha}}
~~{e^{-{\Delta\psi^2_n\over 4\bar{s}}}\over \sinh^2{\Delta\psi_n\over 2}}~~,
\label{3.10}
\end{eqnarray}
where $\Delta\psi_n={A_+\over l}n~,~~A_+=2\pi r_+$.

\noindent
{\bf Rotating black hole ($J\neq 0,~|r_-|\neq 0$)}

When rotation is present we have for the contour integral in (\ref{3.7})
(see (\ref{A5})):
\begin{eqnarray}
&&{\imath\over 4\pi\alpha}
\int_{\Gamma}~{\cot {w\over 2\alpha}~dw\over \sinh^2{\Delta\psi_n\over 2}+
\sin^2({w\over 2}+{\gamma_n\over 2})} \nonumber \\
&&={1\over \alpha}~{\sinh {\Delta\psi_n\over \alpha}\over \sinh\Delta\psi_n}~
{1\over (\sinh^2{\Delta\psi_n\over \alpha}+\sin^2{ [ \gamma_n  ] \over 2\alpha})}-
{1\over (\sinh^2{\Delta\psi_n}+\sin^2{\gamma_n\over 2})}
\label{3.12}
\end{eqnarray}
where $[\gamma ]=\gamma-\pi k,~|[\gamma ]|<\pi$.
Then we obtain for the heat kernel on $B^{\alpha}_3$:
\begin{eqnarray}
&&Tr K_{B_3^{\alpha}}=\left( {V_{B_3^{\alpha}}\over l^3}+{A_+\over l}(2\pi\alpha )
c_2(\alpha )~\bar{s}~ \right){e^{-\mu\bar{s}}\over (4\pi \bar{s})^{3/2}}
\nonumber \\
&&+2\pi {e^{-\mu\bar{s}}\over (4\pi \bar{s})^{3/2}}{A_+\over l}~\bar{s}~
\sum_{n=1}^{\infty}{\sinh {\Delta\psi_n\over \alpha}\over \sinh {\Delta\psi_n}}
~~{e^{-{\Delta\psi^2_n\over 4\bar{s}}}\over (\sinh^2{\Delta\psi_n\over 2\alpha}+\sin^2{[\gamma_n]\over 2\alpha})}~~,
\label{3.13}
\end{eqnarray}
where $\gamma_n=|A_-|n/l$ and $\Delta \psi_n=A_+n/l$. Remarkably,
(\ref{3.13}) has the periodicity 
$\gamma_n\rightarrow \gamma_n+2\pi\alpha$ or equivalently $|A_-|n/l
\rightarrow |A_-|n/l +2\pi\alpha$.

As discussed in Section 2, any result obtained for the Euclidean 
black hole must be analytically continued to  Lorentzian values of the
parameters by means of (\ref{?}).
For the non-rotating black hole this is rather straightforward. 
It simply means that the area $A_+$ of the Euclidean horizon becomes
the area of the horizon in the Lorentzian space-time. 
For a rotating black hole the procedure is more subtle. From (\ref{?}) 
we must  also transform $|A_-|$ which after 
analytic continuation becomes imaginary  ($|A_-|\rightarrow \imath A_-$),
where $A_-$ is area of the lower horizon of the Lorentzian black hole.
Doing this continuation  in the left hand side of the contour integral
(\ref{3.12}) we find that the right hand side becomes
\begin{equation}
\sin^2( {\imath\gamma_n\over 2})=-\sinh^2{\gamma_n\over 2}~,~~
\sin^2({[\imath\gamma_n]\over 2\alpha})=-\sinh^2{\gamma_n\over 2\alpha}~~,
\label{??}
\end{equation}
where $\gamma_n=A_- n/l$. Below we are assuming this kind of substitution 
when we are applying our formulas to the Lorentzian black hole.
We see that after the continuation we lose periodicity with respect
to $\gamma_n$.

It should be noted that there is only
a small group of conical spaces for which the
heat kernel is known explicitly \cite{con}. (The small $s$ expansion 
for the heat kernel on conical spaces has been more widely studied, and 
a rather general result that the coefficients of this expansion contain 
terms (additional to the standard ones) due to the
conical singularity only and are defined on the singular subspace $\Sigma$
has recently been obtained \cite{DF}, \cite{Dowker}.)
However, no black hole geometry among these special cases were known.
In (\ref{3.13}) we have an exact result for a rather 
non-trivial example of a black hole with rotation,
providing us with an exciting possibility to
learn something new about black holes. We consider some of these 
issues in the context of black hole thermodynamics in the next section.

\bigskip
\noindent
{\bf Small $s$ Expansion of the Heat Kernel}

As we can see from Eqs.(\ref{3.10}), (\ref{3.13}) the trace of the heat 
kernel on the conical space $B^{\alpha}_3$ both for the rotating and non-rotating cases has the form
\begin{equation}
Tr K_{B_3^{\alpha}}=\left( {V_{B_3^{\alpha}}\over l^3}+{A_+\over l}(2\pi\alpha )
c_2(\alpha )~\bar{s}~ \right){e^{-\mu\bar{s}}\over (4\pi \bar{s})^{3/2}}~ +~ES~~,
\label{a}
\end{equation}
where $ES$ stands for exponentially small terms which 
behave as $e^{-{1\over s}}$ in the limit $s\rightarrow 0$.
 So, for small $s$ we get the asymptotic formula
\begin{equation}
Tr K_{B_3^{\alpha}}={1\over (4\pi s)^{3/2}} \left(V_{B_3^{\alpha}}+
(-{1\over l^2}V_{B_3^{\alpha}}+{\xi \over l^2}V_{B_3^{\alpha}}+
A_+~(2\pi\alpha )
c_2(\alpha ))~s~ +O(s^2)~\right)
\label{b}
\end{equation}
where $\mu=1-\xi $.

The asymptotic behavior of the heat kernel on various manifolds
is well known and the asymptotic expressions
are derived in terms of  geometrical invariants of the manifold. 
For the operator $(\Box +X)$, where $X$ is some scalar function, 
on a $d$-dimensional  manifold $M^\alpha$ with conical singularity 
whose  angular deficit is $\delta=2\pi (1-\alpha )$  
at the surface $\Sigma$,  
the corresponding expression reads
\begin{equation}
Tr K_{M^{\alpha}}={1\over (4\pi s)^{d/2}} (a_0+a_1~s+O(s^2))~~,
\label{c}
\end{equation}
where
\begin{equation}
a_0=\int_{M^{\alpha}}1~~,~~a_1=\int_{M^{\alpha}}({1\over 6}R+X)~+~(2\pi\alpha )c_2(\alpha )\int_{\Sigma} ~1~~.
\label{d}
\end{equation}
The volume part of the coefficients is standard \cite{BD} while the surface
part in $a_1$ is due to the conical singularity according to \cite{DF}.
One can see that (\ref{b}) exactly reproduces (\ref{c})-(\ref{d})
for operator $(\Box +\xi /l^2 )$ since for the case under consideration 
we have
$R=-6/l^2$.  Note, that in (\ref{a}), (\ref{b}) we do not obtain the usual term
$\int_{\partial M}k$ due to extrinsic curvature $k$ of boundary $\partial M$.
This term does not appear in our case since we calculate the heat kernel
for spaces with boundary lying at infinity where the boundary term
is divergent. But, it would certainly appear if we deal with a boundary staying
at a finite distance. Also, in the expressions (\ref{a}), (\ref{b})
we do not observe a contribution due to extrinsic curvature of the horizon
surface. According to arguments by Dowker \cite{Dowker} such a contribution
to the heat kernel  occurs for generic conical space. 
However, in the case
under consideration the extrinsic curvature of the horizon precisely vanishes.
We observed \cite{MS} the similar phenomenon for charged Kerr black hole
in four dimensions.

\bigskip
\noindent
{\bf Effective action and renormalization}

For the effective action we immedately obtain that
\begin{eqnarray}
&&W_{eff}[B^{\alpha}_3]=-{1\over 2}\int_{\epsilon^2}^{\infty} {ds\over s} Tr K_{B^3_{\alpha}}
\nonumber \\
&&=W_{div}[B^{\alpha}_3]- \sum_{n=1}^{\infty}{1\over 4n}~~{\sinh ({\bar{A}_+
\over \alpha}n)\over \sinh (\bar{A}_+n) }~~ {e^{-\sqrt{\mu}\bar{A}_+ n}\over 
(\sinh^2{\bar{A}_+ n\over 2\alpha}+\sin^2{[|\bar{A}_-|n]\over 2\alpha})}~~,
\label{3.14}
\end{eqnarray}
where the divergent part $W_{div}[B^{\alpha}_3]$ of the effective action
takes the form 
\begin{eqnarray}
&&W_{div}[B^{\alpha}_3]=-{1\over 2}{1\over (4\pi)^{3/2}}\left(V_{B_3^{\alpha}}
\int_{\epsilon^2}^{\infty}
{ds\over s^{5/2}}e^{-\mu s/ l^2}~+~A_+~(2\pi\alpha)~c_2(\alpha )
\int_{\epsilon^2}^{\infty}
{ds\over s^{3/2}}e^{-\mu s/ l^2}~ \right) \nonumber \\
&&=-{1\over (4\pi)^{3/2}}~[V_{B_3^{\alpha}}~ ({1\over 3\epsilon^3}-{
\mu\over l^2\epsilon}+{2\over 3}{\mu^{3/2}\over l^3}\sqrt{\pi}+O(\epsilon ))
\nonumber \\
&& +A_+~(2\pi\alpha)~c_2(\alpha )({1\over \epsilon}-{\sqrt{\mu\pi}\over l}+O(\epsilon ))]~~.
\label{3.11'}
\end{eqnarray}
Recall that Eqs.(\ref{3.14}), (\ref{3.11'}) must be analytically continued
by means of (\ref{?}) and (\ref{??}) to deal with the characteristics of the
Lorentzian  black hole.
Note that the rotation parameter $J$  enters the 
UV-infinite part  (\ref{3.14}) only via $A_+$.
The form of (\ref{3.14}) is therefore the same for rotating 
and non-rotating holes.
Similar behavior for an uncharged Kerr black hole
was previously observed in four dimensions \cite{MS}.

The classical gravitational action
$$
W=-{1\over 16\pi G_B}\int_M (R+{2\over l^2})=-{1\over 16\pi G_B}\int_M R
-\lambda_B \int_M~1~~,
$$
where $\lambda_B={1\over 8\pi G_B} {1\over l^2}$.
In the presence of a conical singularity with angular 
deficit $\delta=2\pi (1-\alpha )$ on a  surface of area $A_+$ 
this has the form
\begin{equation}
W=-{1\over 16\pi G_B}\int_{M^\alpha} R-{1\over 4 G_B}A_+(1-\alpha )
-\lambda_B \int_{M^\alpha}~1~~.
\label{3.21}
\end{equation}
The ${1\over \epsilon}$ and ${1\over \epsilon^3}$ UV-divergences of the
effective action (\ref{3.14})-(\ref{3.11'}) for $\alpha=1$ (regular manifold 
without conical singularities) are known to be absorbed in the 
renormalization of respectively the bare Newton
constant $G_B$ and cosmological constant $\lambda_B$ of the 
classical action. As was pointed out in \cite{SS} and \cite{FS} the 
divergences of the effective action that are of first order with respect 
to $(1-\alpha )$ are automatically removed by the same
renormalization of Newton's constant $G_B$ in the classical action 
(\ref{3.21}). This statement is important in the context of the 
renormalization of UV-divergences of the black hole entropy. Its validity 
in the case under consideration can be easily demonstrated if we note 
that $(2\pi\alpha )c_2 (\alpha )={2\over 3}\pi (1-\alpha )
+O((1-\alpha )^2)$ and define the renormalized quantities $G_{ren}$ 
and $\lambda_{ren}$ as follows:
\begin{equation}
{1\over 16\pi G_{ren}}={1\over 16\pi G_B}+{1\over 12}~
{1\over (4\pi)^{3/2}}~\int_{\epsilon^2}^{\infty}{ds\over s^{3/2}}e^{-\mu s/ l^2}
\label{3.22}
\end{equation}
and
\begin{equation}
\lambda_{ren}=\lambda_B+{1\over 2}~
{1\over (4\pi)^{3/2}}~\left(\int_{\epsilon^2}^{\infty} {ds\over s^{5/2}}e^{-\mu s/ l^2}+
l^2\int_{\epsilon^2}^{\infty} {ds\over s^{3/2}}e^{-\mu s/ l^2} \right)~~.
\label{3.23}
\end{equation}
Then all the divergences in (\ref{3.11'}) which are up to order
$(1-\alpha )$ are renormalized by (\ref{3.22}), (\ref{3.23}). The 
renormalization of terms $\sim O((1-\alpha )^2)$ requires in principle 
the introduction of some new
counterterms. However they are irrelevant for black
hole entropy.

The prescription (\ref{3.22}), (\ref{3.23}) includes in part some
UV-finite renormalization. This is in order that
$G_{ren}$ and $\lambda_{ren}$ be treated as macroscopically measurable 
constants. Note also that the relation between the bare 
constants: $\lambda_B G_B={1\over 8\pi}{1\over l^2}$ is no longer valid 
for the renormalized quantities (\ref{3.22})-(\ref{3.23}). 

\bigskip

\section{Entropy}
\setcounter{equation}0

A consideration of the conical singularity at the horizon for the 
Euclidean black hole is a convenient way to obtain the thermodynamic
quantities of the hole. Geometrically, the angular deficit
$\delta =2\pi (1-\alpha ),~\alpha={\beta\over \beta_H}$ appears when 
we close the Euclidean time coordinate with an arbitrary period 
$2\pi\beta$. Physically it means that we consider the statistical 
ensemble containing a black hole at a temperature 
$T=(2\pi\beta )^{-1}$ different from the Hawking value 
$T_H=(2\pi\beta_H )^{-1}$. The state of the system at the
Hawking temperature is the equilibrium state corresponding to the 
extremum of the free energy \cite{FWS}. The entropy of the black 
hole appears in 
this approach as the result of a small deviation from equilibrium. 
Therefore in some sense the entropy is an off-shell quantity. 
If $W[\alpha ]$ is the action calculated for arbitrary angular 
deficit $\delta$ at the horizon we get 
\begin{equation}
S=(\alpha\partial_\alpha-1)W[\alpha ]|_{\alpha=1}~~.
\label{5.1}
\end{equation}
for the black hole entropy.
Applying this formula to the classical gravitational action (\ref{3.21})
we obtain the classical Bekenstein-Hawking entropy:
\begin{equation}
S_{BH}={A_+\over 4G_B}~~.
\label{5.2}
\end{equation}
Applying (\ref{5.1}) to the (renormalized) quantum action $W+W_{eff}$
(\ref{3.14}), (\ref{3.21})
we obtain  the (renormalized) quantum entropy of black hole:
\begin{eqnarray}
&&S={A_+\over 4G}+\sum_{n=1}^\infty
s_n~~, \nonumber \\
&&s_n={1\over 2n}{e^{-\sqrt{\mu}\bar{A}_+ n}\over (\cosh \bar{A}_+n-\cosh \bar{A}_-n)}
(1+\bar{A}_+ n \coth \bar{A}_+ n \nonumber \\
&&- {(\bar{A}_+n \sinh \bar{A}_+ n -
\bar{A}_-n \sinh \bar{A}_-n )\over (\cosh \bar{A}_+n-\cosh \bar{A}_-n)}
)~~,
\label{5.3}
\end{eqnarray}
where $G\equiv G_{ren}$ is the renormalized Newton constant. We already
have done the analytic continuation (\ref{??}) in (\ref{5.3}) in
order to deal with the characteristics of the Lorentzian  black hole.
The second term in the right hand side of (\ref{5.3}) can be 
considered  to be the one-loop
quantum (UV-finite) correction to the classical entropy of black hole.

Since ${A_-\over A_+}=k<1$, $s_n$ is a non-negative quantity which 
monotonically decreases with $n$ and
has asymptotes:
\begin{equation}
s_n\rightarrow {1\over 4n}e^{-(1+\sqrt{\mu})\bar{A}_+n}~~if~~A_+\rightarrow \infty
\label{5.4}
\end{equation}
and
\begin{equation}
s_n\rightarrow {1\over 6n}-{\mu\over 6}\bar{A}_+ ~~if~~A_+\rightarrow 0~~.
\label{5.5}
\end{equation}
Note that both asymptotes (\ref{5.4}), (\ref{5.5})
are independent of the parameter $A_-$ characterizing 
the rotation of the hole.

The infinite sum in (\ref{5.3}) can be approximated by integral. We
find that
\begin{equation}
S={A_+\over 4G}+\int_{\bar{A}_+}^\infty
~s(x)~dx~~,
\label{5.6}
\end{equation}
where
\begin{equation}
s(x)={1\over 2x}{e^{-\sqrt{\mu}x}\over (\cosh x-\cosh kx)}
\left(1+x \coth x - {(x\sinh x -
kx \sinh kx )\over (\cosh x-\cosh kx)}
\right)~~.
\label{5.7}
\end{equation}
For large enough $\bar{A}_+\equiv {A_+\over l}>>1$ the integral in (\ref{5.6}) exponentially
goes to zero and we have the classical Bekenstein-Hawking formula for entropy.
On the other hand, for  small $\bar{A}_+$ the integral in (\ref{5.6})
 is logarithmically divergent
so that we have
\begin{equation}
S={A_+\over 4G}+{\sqrt{\mu}\over 6}{A_+\over l}-{1\over 6}\ln {A_+\over l}+
O(({A_+\over l})^2)~~.
\label{5.8}
\end{equation} 
This logarithmic divergence can also be understood by
examining the expression (\ref{5.3}). {}From (\ref{5.5}) it follows that
every $n$-mode gives a finite contribution $s_n={1\over 6n}$ at 
zero $A_+$. Their sum $\sum_{n=1}^\infty {1\over 6n}$, however,
is not convergent since $s_n$ does not decrease fast enough. 
This divergence appears as the logarithmic one in (\ref{5.8}). This 
logarithmic behavior for small $A_+$ is universal,
independent of the constant $\xi$ 
(or $\mu$) in the field operator 
and the area of the inner horizon  ($A_-$) of the black hole. Hence
the rotation parameter $J$ enters (\ref{5.8}) only via the area $A_+$ 
of the larger horizon. It should be note that similar 
logarithmic behavior was previously observed in various models both 
in two \cite{SS}, \cite{FWS}
and four \cite{F1}, \cite{Zas} dimensions. Remarkably, it appears in 
the three dimensional model as the result of an
explicit one-loop calculation. 

The first quantum correction to the 
Bekenstein-Hawking entropy due to quantization of the three-dimensional
gravity itself was calculated in \cite{rev2} and was shown to be proportional
to area $A_+$ of outer horizon.

\section{Concluding Remarks}
\setcounter{equation}0

Our computation of the quantum-corrected entropy (\ref{5.6}) 
of the BTZ black hole
has yielded the interesting result that the entropy is not proportional to
the outer horizon area ({\it i.e.} circumference) $A_+$, but instead
develops a minimum for sufficiently small $A_+$.
(The plot of the entropy  as function of area $A_+$ for non-rotating case
is represented in Fig.2.) This minimum is
a solution to the equation
\begin{equation}
{l\over 4G}=s({A_{+  min}\over l})~~.
\label{5.9}
\end{equation}
The constants $G$ and $l$ determine two different scales in the theory. 
The former determines the strength of the gravitational interaction.
The distance $l_{pl}\sim G$ can be interpreted as the Planck scale in this 
theory. It determines the microscopic behavior of quantum 
gravitational fluctuations. 
On the other hand, the constant $l$ (related to curvature via $R=-6/l^2$)
can be interpreted as radius of the Universe that contains the black hole. 
So $l$ is a large distance (cosmological) scale.

Regardless of the relative sizes of $G$ and $l$, the entropy is always
minimized for $A_+ \leq G$.  
If we assume $G<<l$ then (\ref{5.9}) is solved as $A_{+min}={2\over 3}G$, 
However if $G >> l$, then (\ref{5.9}) becomes (for $\mu=0$, say)
$A_+/G \simeq e^{-A_+/l} < 1$. In either case,
the minimum of the entropy occurs for a hole whose horizon 
area is of the order of the
Planck length $r_{+}\sim l_{pl}$. In the process of evaporation the 
horizon area of a hole
typically shrinks. The evaporation is expected to stop when the
black hole takes the minimum entropy configuration. In our case it is the
configuration with horizon area $A_+=A_{+ min}$.  Presumably it has zero 
temperature and its geometry is a reminscent of an extremal black hole.
However at present we cannot definitively conclude this
since our considerations do not take into account quantum back reaction
effects. These effects are supposed to drastically  change the geometry at
a distance $r\sim l_{pl}$. Therefore the minimum entropy configuration is 
likely to have little in common with the classical black hole 
configuration described in Section 2. 
Further investigation of this issue will necessitate
taking the back reaction into  account.

\section*{Acknowledgements}
This work was supported by the Natural Sciences and Engineering Research
Council of Canada and by a NATO Science Fellowship.

\newpage
{\appendix \noindent{\large \bf Appendix: Contour integrals}}\\
\def\theequation{A.\arabic{equation}}
\setcounter{equation}0

Consider the integral
\begin{eqnarray}
&&I={1\over 2\pi \imath}\int_\Gamma f(w)~dw~~, \nonumber \\
&&f(w)=\cot {w\over 2\alpha}~{1\over a^2+\sin^2(
{w+\gamma\over 2})}~~,
\label{A1}
\end{eqnarray}
where Im $a=$ Im $ \gamma=0$ and the 
 contour $\Gamma$ in (\ref{A1}) consists of two vertical
lines, going from $(-\pi+\imath \infty )$ to $(-\pi-\imath \infty )$
and from $(\pi-\imath \infty )$ to $(\pi+\imath \infty )$ and
intersecting the real axis between the poles of the 
$\cot {w\over 2\alpha}$: $-2\pi\alpha,~0$ and $0,~+2\pi\alpha$ 
respectively.

The integration in (\ref{A1}) is carried out by calculating the residues
of the function $f(w)$. Let us assume that $-\pi <\gamma < \pi$ 
($|\gamma|<\pi$). Then the function $f(w)$ has the following poles
and residues:

$a).$ $w=w_0=0$, 

$$Res~ f(w_0)={2\alpha\over {a^2+\sin^2{\gamma\over 2}}}$$

$b).$ $w=w_{\pm}=-\gamma\pm 2\imath A$, 

$$Res~f(w_\pm)={2\imath \over \sinh 2A}~
\left( \pm {\tan {\gamma \over 2\alpha}\over \cosh^2{A\over \alpha}}+
{\imath \tanh {A\over \alpha}\over \cos^2{\gamma\over 2\alpha}} \right)
\left(\tanh^2{A\over \alpha}+\tan^2{\gamma\over 2\alpha}\right)^{-1}~~,$$
where we have introduced $A$ related with $a$ as follows: $\sinh A=a$.

Then the integral (\ref{A1}) reads
\begin{eqnarray}
&&I=\sum_{w=w_0, w_+,w_-} ~Res~f(w)\nonumber \\
&&=\left({2\alpha\over \sinh^2 A+\sin^2{\gamma\over 2}}-{\sinh {2A\over \alpha}
\over \sinh {2A}}~{2\over ( \sinh^2 {A\over \alpha}+\sin^2{\gamma\over 
2\alpha})} \right)~~.
\label{A2}
\end{eqnarray}
Two special cases of this expression are worth noting. If $\gamma=0$
we get from (\ref{A2}):
\begin{equation}
I={2\over \sinh^2A}(\alpha-{\tan A\over \tan {A\over \alpha}})
\label{A3}
\end{equation}
and if $\gamma=a=0$ we get (see also \cite{DF})
\begin{equation}
I={2\over 3}(\alpha-{1\over \alpha})\equiv 
-4\alpha c_2(\alpha )
\label{A4}
\end{equation}

If $\pi<|\gamma |<2\pi$ the structure of the pole at $w=w_0=0$
remains the same as above while the other poles lying in the region
$-\pi <w<\pi$ are $w_\pm=2\pi-\gamma\pm 2\imath A$. 
Hence the corresponding residue
takes the same form $b)$, with the replacment 
$\gamma\rightarrow2\pi-\gamma$. Next, taking $\gamma$ to be arbitrary,
define $[\gamma ]$ as follows:
$|\gamma |=\pi k+ [\gamma ]~,[\gamma ]<\pi$.
Then for arbitrary $\gamma$ the integral (\ref{A1}) reads
\begin{equation}
I=2\left({\alpha\over \sinh^2 A+\sin^2{\gamma\over 2}}-{\sinh {2A\over \alpha}
\over \sinh {2A}}~{1\over ( \sinh^2 {A\over \alpha}+\sin^2{[\gamma ]\over 
2\alpha})} \right)~~.
\label{A5}
\end{equation}

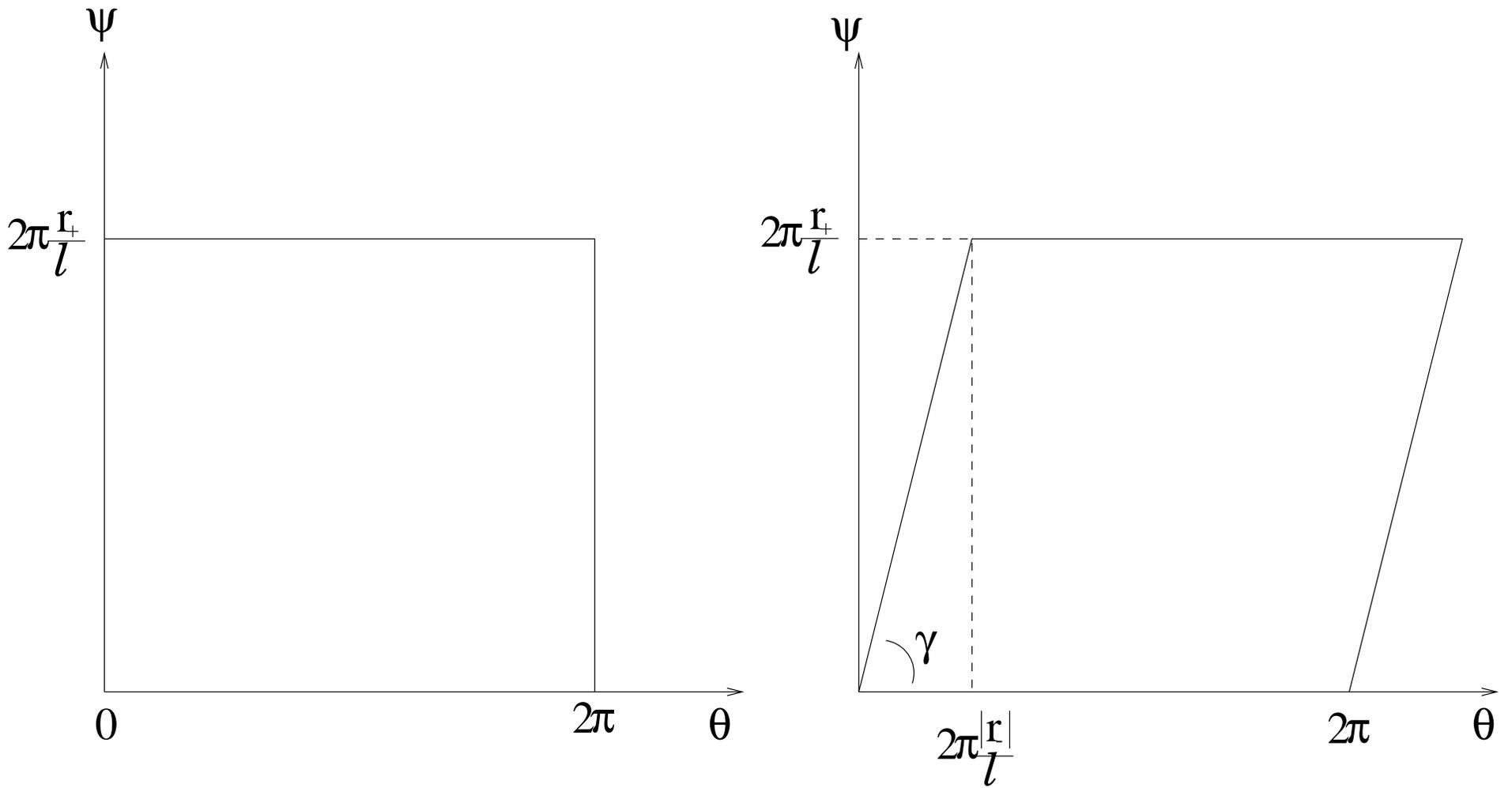
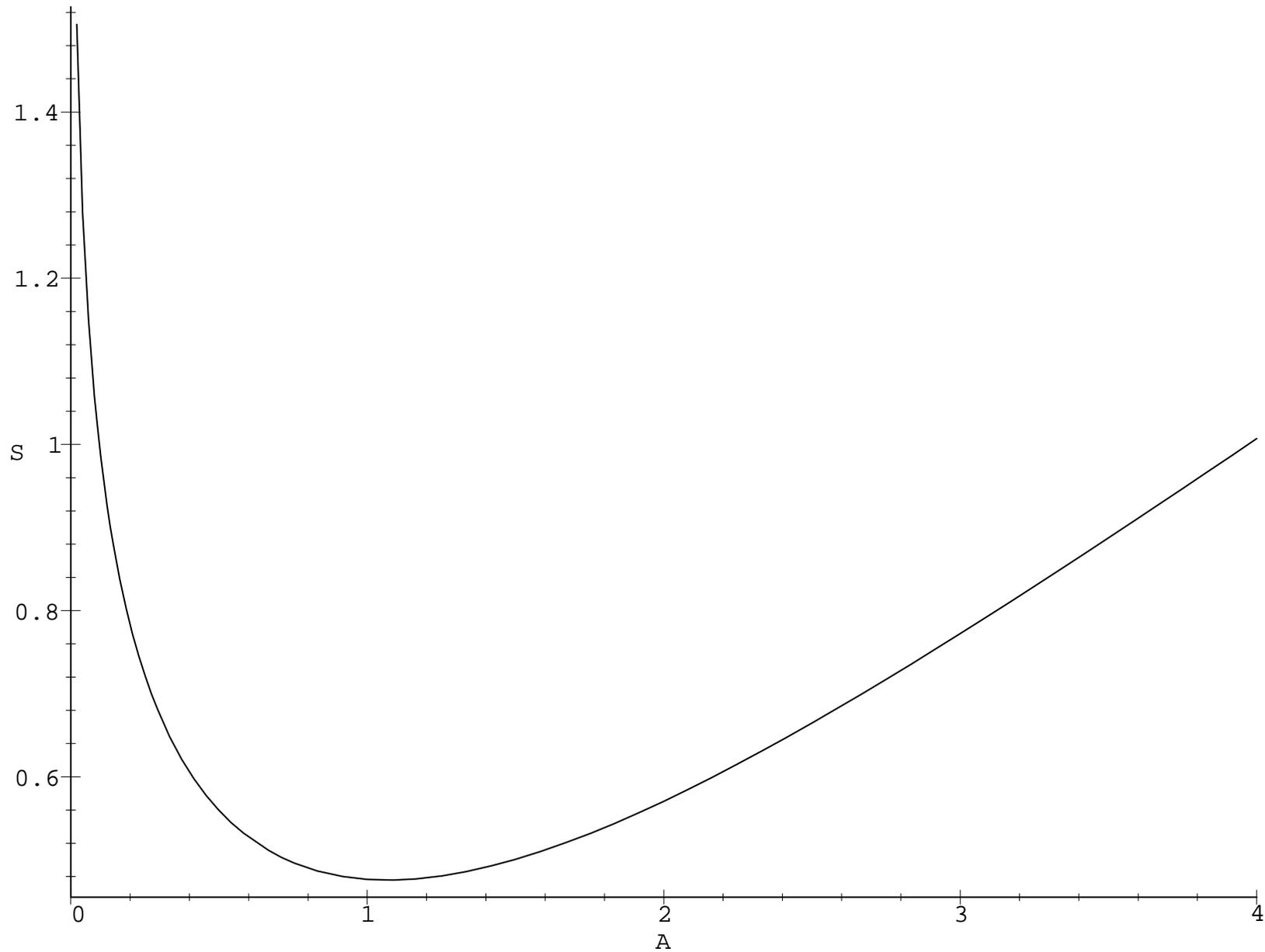
\end{document}